\documentclass[aps,prl,twocolumn,groupedaddress]{revtex4}
\usepackage{amsfonts}
\usepackage{amsmath}
\usepackage{graphicx,epsfig}
\usepackage{color}
\usepackage[pdftex,colorlinks=false,bookmarks]{hyperref}
\usepackage{mathptm}
\usepackage{amssymb}
\usepackage{graphicx}

\input{Qcircuit}
\begin{document}

\title{Anyonic interferometry and protected memories in atomic spin lattices}
\author{Liang~Jiang$^{1}$}
\author{Gavin~K.~Brennen$^{2}$}
\author{Alexey~V.~Gorshkov$^{1}$}
\author{Klemens~Hammerer$^{2}$}
\author{Mohammad~Hafezi$^{1}$}
\author{Eugene~Demler$^{1}$}
\author{Mikhail~D.~Lukin$^{1}$}
\author{Peter~Zoller$^{2}$}
\affiliation{$^{1}$Physics Department, Harvard University, Cambridge, MA 02138, USA}
\affiliation{$^{2}$Institute for Theoretical Physics, University of Innsbruck, and
Institute for Quantum Optics and Quantum Information of the Austrian Academy
of Science, 6020 Innsbruck, Austria}
\date{\today }

\begin{abstract}
\end{abstract}

\maketitle

\textbf{Strongly correlated quantum systems can exhibit exotic behavior
called topological order which is characterized by non-local correlations
that depend on the system topology. Such systems can exhibit remarkable
phenomena such as quasi-particles with anyonic statistics and have been
proposed as candidates for naturally fault-tolerant quantum computation.
Despite these remarkable properties, anyons have never been observed in
nature directly. Here we describe how to unambiguously detect and
characterize such states in recently proposed spin lattice realizations
using ultra-cold atoms or molecules trapped in an optical lattice. We
propose an experimentally feasible technique to access non-local degrees of
freedom by performing global operations on trapped spins mediated by an
optical cavity mode. We show how to reliably read and write topologically
protected quantum memory using an atomic or photonic qubit. Furthermore, our
technique can be used to probe statistics and dynamics of anyonic
excitations.}

By definition, topologically ordered states \cite{Wen04} cannot be
distinguished by local observables, i.e. there is no local order parameter.
They can arise as ground states of certain Hamiltonians which have
topological degeneracy and which provide robustness against noise and
quasi-local perturbations. These properties of such systems are attractive
for quantum memories. However, the local indistinguishability makes
measuring and manipulating the topological states difficult because they are
only coupled by global operations. One way to access this information is to
measure properties of the low lying particle-like excitations. In two
dimensions, the quasi-particles act like punctures in a surface which can
have anyonic statistics and the topological properties are probed by
braiding different particle types around each other. The existence of anyons
also implies a topological degeneracy \cite{Einarsson90}. Quantum Hall
fluids at certain filling fractions are believed to be topologically
protected and there is a vigorous experimental effort to verify anyonic
statistics in these systems \cite{DasSarma07}. A standard approach is to
perform some kind of interferometry where one looks for non-trivial action
on the fusion state space upon braiding. This is manifested as the evolution
of a non-trivial statistical phase in the abelian case, or a change in the
amplitude of the participating states in the non-abelian case. Some
experimental evidence consistent with observation of abelian anyonic
statistics in a $\nu =2/5$ filled Quantum Hall state has been reported \cite%
{Camino05} but an unambiguous detection of anyons is still considered an
open issue \cite{Rosenow07}.

Spin lattice Hamiltonians can also exhibit topological order and such
Hamiltonians can be built with atoms \cite{Duan03} or molecules \cite%
{Micheli06} trapped in an optical lattice. A significant advantage of using
atomic systems is that the microscopic physics is well known and there are
established techniques for coherent control and measurement. Suggestions
have been made for how one would design anyonic interferometers in these
systems by using local spin operations to guide excitation along braiding
paths \cite{Brennen07,Pachos07,Zhang06}.

We here present a new approach that directly measures topological degeneracy
and anyonic statistics using global operations. The technique involves
coupling between a probe qubit (single ancilla spin qubit or optical mode)
and topologically ordered atomic spins in an optical lattice. A many body
interaction between spins is mediated by coupling to a common bosonic mode
of the radiation field via techniques of cavity QED \cite%
{Mabuchi02,Gupta07,Colombe07,Brennecke07} or, alternatively, via a common
phonon mode in ion traps \cite{Cirac95}. Our approach avoids localizing and
guiding excitations while enabling the measurement of the statistical phase
associated with arbitrary braiding paths.

We also note that recent experiments have demonstrated braiding operations
on small networks of non-interacting qubits encoded in photon polarization
\cite{Lu07,Pachos07b}, which generates a simulation of anyonic
interferometry \cite{Han07}. However, since the background Hamiltonian
vanishes in such systems, they are not protected from noise and the particle
interpretation of the \textquotedblleft excitations" is ambiguous. In
contrast, the technique developed here allows one to probe directly dynamic
evolution of anyonic quasi-particles within the parent Hamiltonian.
In addition, our mechanism can be used to perform reading and writing of
qubits initially encoded in light or atoms into topological memory, which
may be useful for offline storage during a computation and for applications
in long distance quantum communication \cite{Briegel98,JTKL07}.

\section{ATOMIC AND MOLECULAR SPIN LATTICES IN OPTICAL CAVITIES}

\begin{figure}[tbp]
\begin{center}
\includegraphics[width=8.7cm]{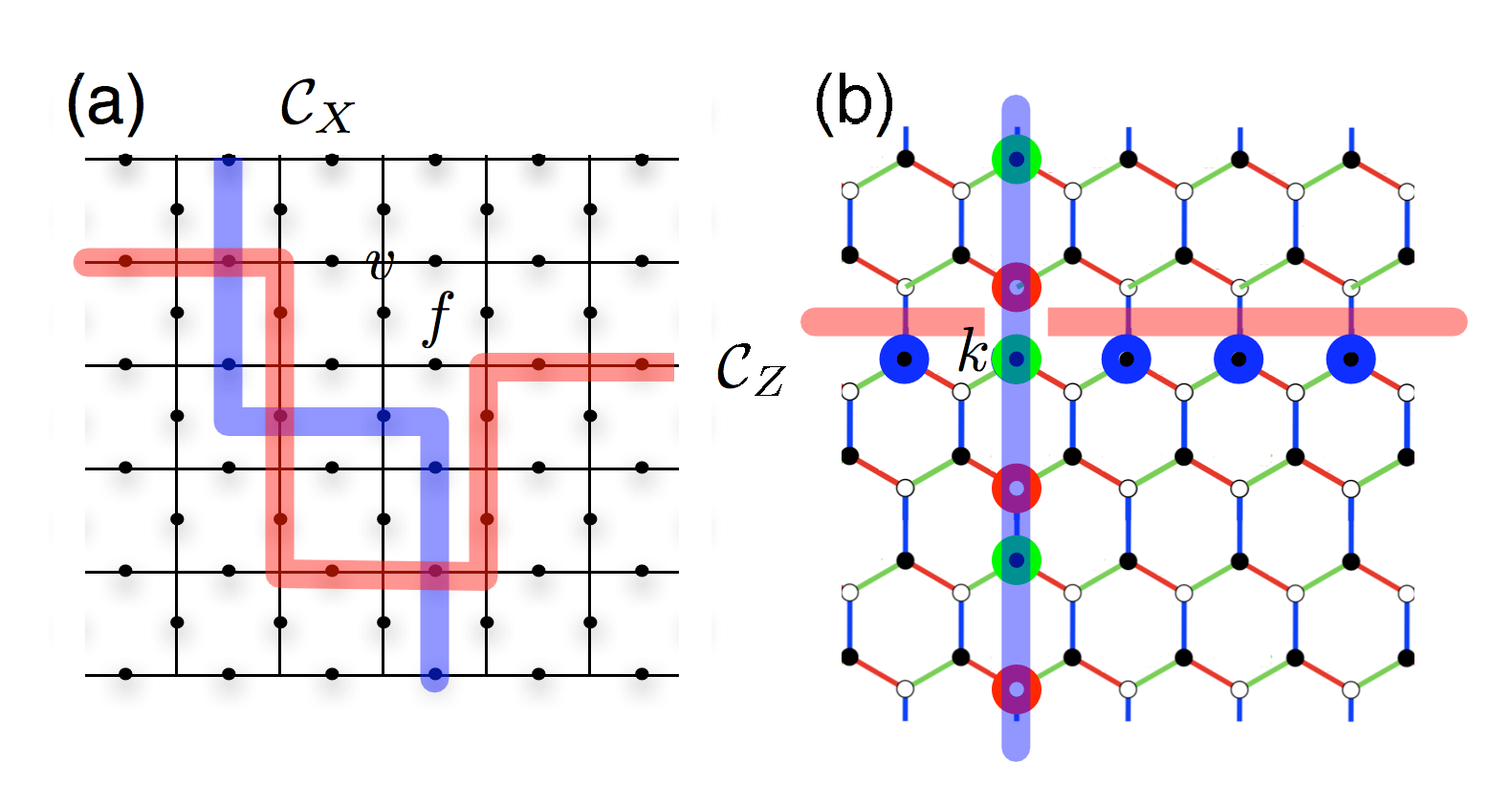}
\end{center}
\caption[fig:lat]{Generators for the encoded qubits. (a) A planar code which
encodes one logical qubit in the ground states. There is a spin-1/2 particle
(filled dot) for each edge of the lattice. The interactions of the local
Hamiltonian $H_{\mathrm{surf}}$ are along edges that bound a face $f$, and
edges that meet at a vertex $v$. The strings $\mathcal{C}_{X,Z}$ indicate
paths of products of $\protect\sigma ^{x,z}$ operators that are logical
operators on the code. (b) A nearest neighbor local Hamiltonian $H_{\mathrm{%
hcb}}$ on the honeycomb lattice, with a spin-1/2 particle for each lattice
site. The (green, red, blue) edges represent interactions of type ($\protect%
\sigma ^{x}\protect\sigma ^{x}$ ,$\protect\sigma ^{y}\protect\sigma ^{y}$ ,$%
\protect\sigma ^{z}\protect\sigma ^{z}$). In the limit that the interactions
along the blue links are much stronger than those along the other links, the
ground subspace has a gapped $\mathbb{Z}_{2}$ topological phase \protect\cite%
{Kitaev06}. Physical $(\protect\sigma ^{x},\protect\sigma ^{y},\protect%
\sigma ^{z})$ spin operations as part of the strings $\mathcal{C}_{X,Z}$ are
indicated by bold (green, red, blue) circles around the spins. At qubit $k$,
the string crossing, the operation is $\protect\sigma _{k}^{x}\protect\sigma %
_{k}^{z}$.}
\label{fig:lat}
\end{figure}

We focus on physical systems in which a two-dimensional optical lattice is
placed within a high-finesse optical cavity as illustrated in Fig.~\ref%
{fig:TopoMem1}a. To be specific, we consider the 2D square lattice
Hamiltonian introduced by Kitaev \cite{Kitaev03} where each edge of the
lattice represents a spin-1/2 particle (see Fig.~\ref{fig:lat}a). Each
vertex $v$ or each face $f$ is associated with an operator $H_{v}=\Pi_{j\in
\mathrm{{star}(v)}}\sigma_{j}^{x}$ or $H_{f}=\Pi_{j\in\partial f}\sigma
_{j}^{z}$. These operators collide on an even number of edges and hence
mutually commute. We seek to encode in the $+1$ coeigenspace of these local
stabilizers by choosing the so-called \emph{surface-code Hamiltonian}:
\begin{equation}
H_{\mathrm{surf}}=-U\sum_{v}H_{v}-J\sum_{f}H_{f}.
\end{equation}
$(U,J>0)$. The ground states of $H_{\mathrm{surf}}$ have a degeneracy $%
\mathrm{dim}\mathcal{H}_{gr}=2^{2g+h}$ where $g$ is the genus of the surface
and $h$ is the number of holes \cite{Dennis02}. Designing lattices with
genus $g>0$, such as the surface of a torus, is challenging, but it is
possible to create several holes ($h>0$) in a planar lattice by, for
instance, deactivating regions of the lattice with focused far detuned
lasers. Alternatively, the planar code with specific boundary as shown in
Fig.~\ref{fig:lat}a provides a ground state degeneracy of $2$. The logical
states are coupled by the operators: $\tilde{Z}=\Pi_{j\in\mathcal{C}%
_{Z}}\sigma_{j}^{z}$ and $\tilde{X}=\Pi_{j\in\mathcal{C}_{X}}\sigma_{j}^{x}$
where the configurations $C_{Z}(C_{X})$ are strings on the lattice (dual
lattice) as illustrated in Fig.~\ref{fig:lat}a.

There are several experimental proposals to implement the spin lattice
Hamiltonians with topological order. For example, Kitaev's honeycomb lattice
Hamiltonian $H_{\mathrm{hcb}}$ (see Fig.~\ref{fig:lat}b)\cite{Kitaev06} can
be designed in optical lattices with ultracold atoms using controlled spin
exchange interactions \cite{Duan03}, or with molecules using microwave
induced dipole-dipole interactions \cite{Micheli06}. With an appropriate
choice of coupling parameters \cite{Kitaev06}, the honeycomb lattice
Hamiltonian has a gapped abelian phase with a low energy effective
Hamiltonian locally equivalent to $H_{\mathrm{surf}}$. In the following, we
will assume the system interacts via $H_{\mathrm{surf}}$, but our results
are also applicable to other spin lattice Hamiltonians.


We now consider how to implement the global operations for the spin lattice
system. In particular, we are interested in a specific type of global
operation: products of Pauli operators on a set of spins whose corresponding
edges in the lattice form a connected string. Such global operators are
called \emph{string operators}. For example, the generators for the encoded
qubits ($\tilde{Z}$ and $\tilde{X}$) are string operators (see Fig.~\ref%
{fig:lat}a). All string operators are equivalent to $S_{\mathcal{C}%
}^{z}=\prod_{j\in C}\sigma_{j}^{z}$ up to local single spin rotations, where
$\mathcal{C}$ is the set of selected spins. For example, $S_{\mathcal{C}%
}^{x}=\prod_{j\in C}\sigma_{j}^{x}=\prod_{j\in C}H_{j}\sigma_{j}^{z}H_{j}$
where $H_{j}$ is the Hadamard rotation for the $j$th spin.

\begin{figure}[tbp]
\begin{center}
\includegraphics[width=8.7cm]{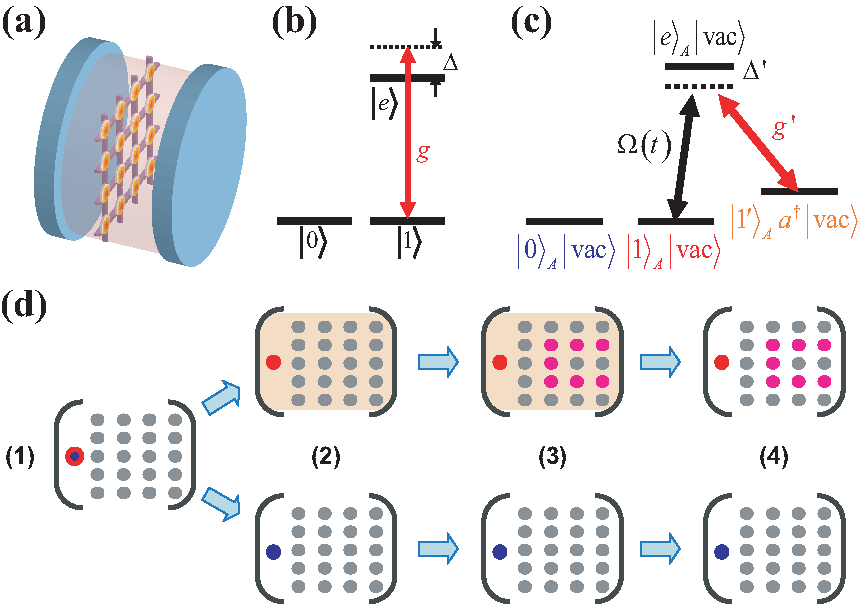}
\end{center}
\caption[fig:TopoMem1]{(a) Inside a cavity, an optical lattice carries spins
for topological memory, with individual spin addressability \protect\cite%
{Cho07,Gorshkov07}. (b) The energy levels of a selected memory spin
interacting dispersively with the cavity mode, which implements the QND\
Hamiltonian of equation~(\protect\ref{QNDHamiltonian}). The coupling
coefficient is $\protect\chi =g^{2}/\Delta $, with single-photon Rabi
frequency $g$ and detuning $\Delta $. (c) The energy levels of the ancilla
spin and the cavity mode for the single photon approach. The control laser $%
\Omega \left( t\right) $ connects the states $\left\vert 1\right\rangle
_{A}\otimes \left\vert \mathrm{vac}\right\rangle $ and $\left\vert 1^{\prime
}\right\rangle _{A}\otimes a^{\dag }\left\vert \mathrm{vac}\right\rangle $,
and enables coherent creation and absorption of a cavity photon conditioned
on the ancilla spin. (d) Cartoon illustration of the procedure for the
implementation of single-photon approach for controlled-string operations:
(1) Initialize the ancilla spin (the left highlighted spin) in a
superposition state $\protect\alpha \left\vert 0\right\rangle _{A}+\protect%
\beta \left\vert 1\right\rangle _{A}$ (blue for $\left\vert 0\right\rangle
_{A}$ and red for $\left\vert 1\right\rangle _{A}$), with no photon in the
cavity and state $\left\vert \protect\psi \right\rangle _{S}$ for the
topological memory. (2) Coherently create a cavity photon (orange shade) for
ancilla spin state $\left\vert 1\right\rangle _{A}$ (upper branch); no
photon is created for ancilla spin state $\left\vert 0\right\rangle _{A}$
(lower branch). (3) Switch on the interaction between the cavity photon and
the selected spins. If there is a cavity photon (orange shade), a
non-trivial evolution $S_{\mathcal{C}}^{z}$ (pink dots) is implemented. (4)
Turn off the interaction and coherently absorb the cavity photon into the
ancilla spin. Finally the state $\protect\alpha \left\vert 0\right\rangle
_{A}\otimes \left\vert \protect\psi \right\rangle _{S}+\protect\beta %
\left\vert 1\right\rangle _{A}\otimes S_{\mathcal{C}}^{z}\left\vert \protect%
\psi \right\rangle _{S}$ is prepared. }
\label{fig:TopoMem1}
\end{figure}

In our setup, the topological memory consists of a spin lattice of trapped
atoms or molecules inside an optical cavity as illustrated in Fig.~\ref%
{fig:TopoMem1}a. The off-resonant interaction between the common cavity mode
and selected spins is described by the quantum non-demolition (QND)
Hamiltonian \cite{WM94,Scully97}:%
\begin{equation}
H=\chi a^{\dag}a\sum_{j\in C}\sigma_{j}^{z},  \label{QNDHamiltonian}
\end{equation}
which is achieved by choosing the cavity mode with a large detuning $\Delta$
from a spin-dependent optical transition as shown in Fig.~\ref{fig:TopoMem1}%
b. The coupling strength is $\chi=g^{2}/2\Delta$ where $g$ is the
single-photon Rabi frequency for the cavity mode. The QND Hamiltonian
preserves the photon number $n_{a}=a^{\dag}a$ of the cavity mode. In
addition, the cavity mode also interacts with an ancilla spin that probes
anyonic statistics associated with quasi-particles.

Similar to the previous schemes \cite{Brennen07,Zhang06} to measure anyonic
statistics, we assume selective addressing of spins in the lattice so that
we can perform single spin rotations as well as switch on/off the coupling
between the cavity mode and the spins. Such selective manipulation can be
achieved using addressing lasers with shaped intensity profiles \cite%
{Cho07,Gorshkov07}. The key new ingredient, however, is that we use the
common cavity mode to mediate global string operators. In this way, we avoid
problems involving maintaining adiabaticity and localization while braiding
quasi-particles. And most importantly, we are able to achieve \emph{%
controlled-string operations} ( $\Lambda\left[ S_{\mathcal{C}}^{x,z} \right]
$) for an arbitrary string $\mathcal{C}$.

The idea of controlled-string operations can be illustrated by considering a
situation when the cavity mode is first prepared in some superposition of
zero and one photon states. Within this subspace, the evolution of the QND\
Hamiltonian for interaction time $\tau=\pi/ 2\chi$ yields
\begin{align}
U & =\exp\left[ -iH\tau\right] =\left[ \left( -i\right) ^{N_{C}}\prod_{j\in
C}\sigma_{j}^{z}\right] ^{n_{a}}  \label{eq:ParityMeasPhoton} \\
& =\left\{
\begin{tabular}{cc}
$\mathbf{I}$ & for $n_{a}=0$ \\
$\left( -i\right) ^{N_{C}}\prod_{j\in C}\sigma_{j}^{z}$ & for $n_{a}=1$%
\end{tabular}
\ \ \ \ \ \ \ \ \ \ \ \ \ \ \right. ,  \notag
\end{align}
where $N_{\mathcal{C}}$ is the number of elements in $\mathcal{C}$, and the
second equality uses the identity $\exp\left[ -i\frac{\pi}{2}\sigma_{j}^{z}%
\right] =-i\sigma_{j}^{z}$. This unitary evolution will apply the string
operator $S_{\mathcal{C}}^{z}$ to the topological memory, conditioned on one
cavity photon. With such controlled-string operations, we can conveniently
access the topological memory, and build anyonic interferometry to probe
braiding statistics and dynamics of quasi-particles.

In practice, however, it is actually easier to control the ancilla spin
rather than to directly manipulate the photon number state. Therefore, in
the following, we will present two approaches to controlled-string
operations between the ancilla spin and the topological memory.


\section{Controlled-string operations}

The key operation of the single photon approach is the evolution of the QND
interaction described by equation~(\ref{eq:ParityMeasPhoton}). In addition,
the cavity mode interacts with a single ancilla spin using spectroscopically
resolvable energy levels as shown in Fig.~\ref{fig:TopoMem1}c. Starting with
no photon in the cavity mode $\left\vert \mathrm{vac}\right\rangle $ and
ancilla spin in state $\alpha\left\vert 0\right\rangle _{A}+\beta\left\vert
1\right\rangle _{A}$, we can coherently couple the number state of the
cavity mode with the state of the ancilla spin by adiabatically increasing
the Rabi frequency $\Omega\left( t\right) $ of the control laser until it is
much larger than the single-photon Rabi frequency $g^{\prime}$. The
intermediate state is then $\alpha\left\vert 0\right\rangle
_{A}\otimes\left\vert \mathrm{vac}\right\rangle -\beta\left\vert
1^{\prime}\right\rangle _{A}\otimes a^{\dag}\left\vert \mathrm{vac}%
\right\rangle $, having the photon number fully correlated with the ancilla
spin. Applying the QND\ interaction with the intermediate state realizes the
desired controlled-string operation conditioned on the state of the ancilla
spin. Finally, we can reverse the state mapping by adiabatically decreasing
the Rabi frequency, which coherently annihilates the photon of the cavity
mode and restores the ancilla spin to its logical subspace spanned by $%
\left\{ \left\vert 0\right\rangle _{A},\left\vert 1\right\rangle
_{A}\right\} $. Following the procedure summarized in Fig.~\ref{fig:TopoMem1}%
d, we can achieve the controlled-string operation:%
\begin{equation}
\Lambda\left[ S_{\mathcal{C}}^{z}\right] =\left\vert 1\right\rangle
_{A}\left\langle 1\right\vert \otimes S_{\mathcal{C}}^{z}+\left\vert
0\right\rangle _{A}\left\langle 0\right\vert \otimes\mathbf{I.}
\label{eq:CS}
\end{equation}

\begin{figure}[tbp]
\begin{center}
\includegraphics[width=8.7cm]{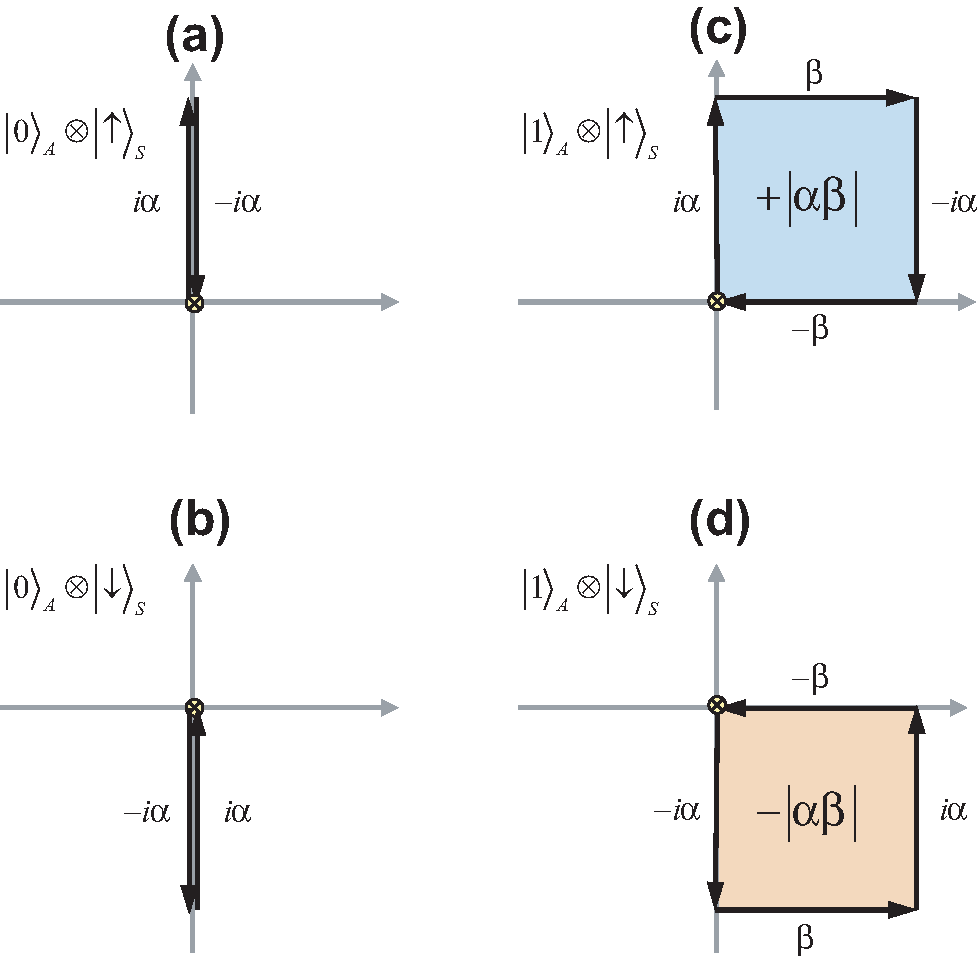}
\end{center}
\caption[fig:GeoPhaseGate]{Phase accumulation for the approach with
geometric phase gate [equation~(\protect\ref{eq:GeoEvolution})]. We use $%
\left\vert \uparrow \right\rangle _{S}$ and $\left\vert \downarrow
\right\rangle _{S}$ to represent $+1$ and $-1$ subspaces of memory spins
associated with the string operator $S_{\mathcal{C}}^{z}$, respectively.
(a)(b) When the ancilla spin is in $\left\vert 0\right\rangle _{A}$ state,
the enclosed area vanishes. When the ancilla spin is in $\left\vert
1\right\rangle _{A}$ state, (c) for the subspace $\left\vert \uparrow
\right\rangle _{S}$ the enclose area is $\left\vert \protect\alpha \protect%
\beta \right\vert $; (d) for the subspace $\left\vert \downarrow
\right\rangle _{S}$ the enclosed area is $-\left\vert \protect\alpha \protect%
\beta \right\vert $. The quantum phase accumulated is twice the area
enclosed.}
\label{fig:GeoPhaseGate}
\end{figure}

The second approach to controlled-string operations is based on the idea of
geometric phase gates \cite{Wang02}. Here, the bosonic field of the cavity
mode starts in a coherent state, rather than a superposition of zero and one
photon states. If our transformation restores the bosonic field to the
initial coherent state, the entire system will accumulate a quantum phase
(geometric phase), which is twice the area enclosed by the trajectory in
phase space of the bosonic field. We activate the geometric phase gate using
an ancilla spin which experiences the QND interaction with the cavity mode
that can be selectively turned on and off \cite{Cho07,Gorshkov07}.
As illustrated in Fig.~\ref{fig:GeoPhaseGate} and detailed in the Methods
section: if the ancilla spin is in state $\left\vert 0\right\rangle _{A}$,
the enclosed area vanishes; if the ancilla spin is in state $\left\vert
1\right\rangle _{A}$, the enclose area has a different sign depending on
whether the topological memory is in $+1$ or $-1$ subspace associated with
the string operator $S_{\mathcal{C}}^{z}$, yielding again equation~(\ref%
{eq:CS}).

Various imperfections such as photon loss and deviation of the QND
interaction can degrade the controlled-string operation. However, we can use
a cavity with high Purcell factor $P$ to suppress photon loss \cite%
{Purcell46,Michler00}, and apply quantum control techniques to suppress the
deviation of the QND interaction to arbitrarily high order \cite%
{Vandersypen04,Brown04}. As derived in the Methods section, the error
probability for controlled-string operation is approximately $\sqrt{N_{%
\mathcal{C}}/P}$, with $N_{\mathcal{C}}$ for the length of the string.

\section{Accessing topological quantum memory}

Controlled-string operations provide an interface between the probe qubit
which features easy access and efficient manipulation, and the topological
memory which provides good storage. To store quantum states we require two
operations: the SWAP$_{\mathrm{in}}$ gate which swaps the state of a probe
qubit $A$ to memory $M$ initialized in $\left\vert \tilde{0}\right\rangle
_{M}$ and the SWAP$_{\mathrm{out}}$ which swaps back to a probe qubit
prepared in $\left\vert 0\right\rangle _{A}$.
\begin{align*}
\mathrm{SWAP}_{\mathrm{in}}& =H_{A}\cdot \Lambda \left[ \tilde{Z}\right]
\cdot H_{A}\cdot \Lambda \left[ \tilde{X}\right] , \\
\mathrm{SWAP}_{\mathrm{out}}& =\Lambda \left[ \tilde{X}\right] \cdot
H_{A}\cdot \Lambda \left[ \tilde{Z}\right] \cdot H_{A},
\end{align*}%
where $H_{A}$ is the Hadamard gate acting on the probe qubit, and $\Lambda %
\left[ \tilde{S}\right] $ represents a controlled-string operation. In
addition, universal rotations of the encoded qubit (generally, arbitrary
unitaries generated by string operators) over the topological memory can be
implemented either by teleportation of quantum gates or by direct geometric
phase gate. (See SOM\ for details.) We remark that the ancilla spin can also
be entangled with another ancilla spin from a different cavity via
probabilistic entanglement generation, and therefore our topological
memories can be used for quantum networks \cite%
{Duan04,Lim06,Benjamin06,JTSL07,JTSL07b}.

To compare the topological memory and unprotected single-spin memory, we
introduce the decoherence rate $q$ for the unprotected spin due to
low-frequency noise. The topological memory can significantly reduce the
decoherence rate by a factor of $\left( \delta h/J\right) ^{N}\ll 1$, where $%
\delta h$ is the magnitude of the noise perturbation on individual spins and
$N$ is the length of the minimal string associated with the generators for
encoded qubits \cite{Kitaev03}. Meanwhile, since the topological memory is
not protected from errors associated with controlled-string operations (with
probability $\sim \sqrt{N/P}$), we should take them into account. Therefore,
in terms of total error probability, the topological memory outperforms the
single-spin memory for storage time $t\gtrsim \frac{1}{q}\sqrt{N/P}$. (See
the Methods section for details).


\section{Anyonic interferometry}

%

We now describe how to use controlled-string operations to extract the
statistical phase acquired when braiding abelian anyons. The definition of
anyonic statistics usually relies on the adiabatic transport of
quasi-particles around each other \cite{Arovas84}, with the required
condition of adiabaticity to keep the system in the same manifold of excited
states and prevent exciting additional degrees of freedom. Note that this
procedure relies explicitly on the existence of the Hamiltonian. This is
fundamentally different from anyonic simulation approaches \cite%
{Han07,Lu07,Pachos07b} not using topological Hamiltonian, which only probe
the non-trivial commutation relations of spin operators and initially
entangled quantum states. However, anyonic statistics is a property of
quasi-particles associated with the Hamiltonian and not just with some
specially prepared initial state.

\begin{figure}[tbp]
\begin{center}
\includegraphics[width=8.7cm]{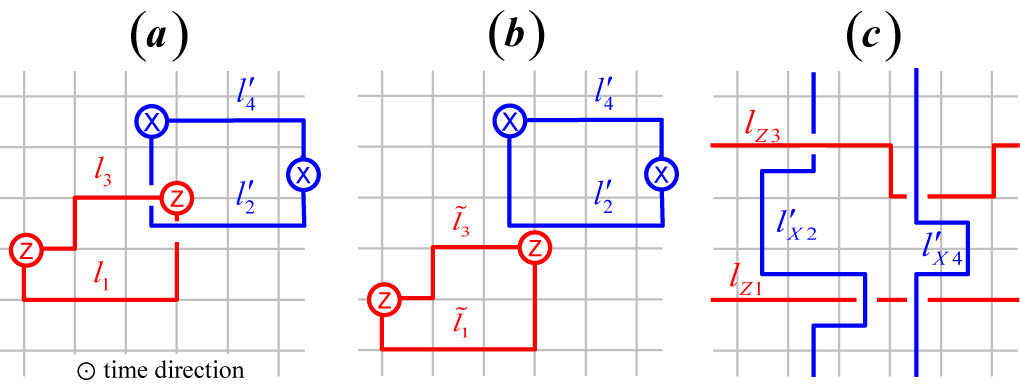}
\end{center}
\caption[fig:AnyonBraiding]{Braiding operations. (a)/(b) We can move
x-particles and z-particles in tangled/untangled loops using string
operators to implement operations with/without braiding of anyons. (c) We
can also apply generators for the encoded qubits to achieve the braiding
operation. The braiding statistics of anyons should be invariant under
non-crossing deformations of the loops \protect\cite{Freedman05}.}
\label{fig:AnyonBraiding}
\end{figure}

For our spin lattice system with $H_{\mathrm{surf}}$, there are two types of
anyons \cite{Kitaev03}: (1) z-particles that live on the vertices of the
lattice and (2) x-particles that live on the face (see Fig.~\ref%
{fig:AnyonBraiding}a,b). These anyons are created in pairs (of the same
type) by string operators: $\left\vert \psi ^{z}\left( l\right)
\right\rangle =S_{l}^{z}\left\vert \xi \right\rangle $ and $\left\vert \psi
^{x}\left( l\right) \right\rangle =S_{l^{\prime }}^{x}\left\vert \xi
\right\rangle $, where $\left\vert \xi \right\rangle $ is some ground state
of the spins, and $S_{l}^{z}=\Pi _{j\in l}\sigma _{j}^{z}$ and $S_{l^{\prime
}}^{x}=\Pi _{j\in l^{\prime }}\sigma _{j}^{x}$ are string operators
associated with string $l$ on the lattice and string $l^{\prime }$ on the
dual lattice, respectively (see Fig.~\ref{fig:AnyonBraiding}). In our
approach, string operators can be used to effectively move quasi-particles
quickly along the string trajectory but without exciting other
quasi-particles. For example, effective motion of quasi-particles
with/without braiding is shown in Fig.~\ref{fig:AnyonBraiding}a,b. This
evolution is described by
\begin{equation}
S_{l_{4}^{\prime }}^{x}U_{t_{3}}S_{l_{3}}^{z}U_{t_{2}}S_{l_{2}^{\prime
}}^{x}U_{t_{1}}S_{l_{1}}^{z}\left\vert \Psi _{initial}\right\rangle
=e^{i\theta _{\mathrm{tot}}}\left\vert \Psi _{initial}\right\rangle ,
\label{braiding3}
\end{equation}%
where we introduce time delays, represented by unitary evolution $U_{t}$,
between string operations. The goal of these delays is to check that the
system stays in the manifold with a fixed number of quasi-particles where
time delays lead to only a trivial dynamical phase. On the other hand, if
the string operator were to create some complicated intermediate states,
time delays would lead to complete decoherence. The total phase $e^{i\theta
_{\mathrm{tot}}}$ includes both the dynamical contribution $e^{i\eta
}=e^{i2J\left( t_{1}+2t_{2}+t_{3}\right) }$ and the statistical contribution
$e^{i\theta }=-1$ (or $+1$) in the presence (or absence) of braiding.
Therefore, we can unambiguously measure the statistical phase if we can
measure $e^{i\theta _{\mathrm{tot}}}$ for both cases.


The following interference experiment can be used to measure the phase $%
e^{i\theta _{\mathrm{tot}}}$. First, we prepare the probe qubit in a
superposition state $(\left\vert 0\right\rangle +\left\vert 1\right\rangle )/%
\sqrt{2}$. We then use controlled-string operations to achieve interference
of the following two possible evolutions: if the probe qubit is in state $%
\left\vert 0\right\rangle $, no operation is applied to the memory spins; if
the probe qubit is in state $\left\vert 1\right\rangle $, the operation $%
S_{l_{4}^{\prime }}^{x}U_{t_{3}}S_{l_{3}}^{z}U_{t_{2}}S_{l_{2}^{\prime
}}^{x}U_{t_{1}}S_{l_{1}}^{z}$ is applied to the topological memory, which
picks up the extra phase $e^{i\theta _{\mathrm{tot}}}$ we want to measure.
After the controlled-string operations, the probe qubit will be in state $%
(\left\vert 0\right\rangle +e^{i\theta _{\mathrm{tot}}}\left\vert
1\right\rangle )/\sqrt{2}$. Finally, we project the probe qubit to the basis
of $\left\vert \xi _{\pm }\right\rangle \equiv (\left\vert 0\right\rangle
\pm e^{i\phi }\left\vert 1\right\rangle )/\sqrt{2}$ with $\phi \in \lbrack
0,2\pi )$, and measure the operator $\sigma _{\phi }\equiv \left\vert \xi
_{+}\right\rangle \left\langle \xi _{+}\right\vert -\left\vert \xi
_{-}\right\rangle \left\langle \xi _{-}\right\vert $. The measurement of $%
\left\langle \sigma _{\phi }\right\rangle $ v.s. $\phi $ should have fringes
with perfect contrast and a maximum shifted by $\phi =\theta _{\mathrm{tot}}$%
. In fact this scheme can be used to measure abelian statistics for an
arbitrary finite cyclic group as described in the Methods section.

It is crucial to verify that the outcome of the anyonic interferometry is
invariant under repeated experiments with deformed string operators (see
Fig.~\ref{fig:AnyonBraiding}) \cite{Freedman05}. For example, the two ground
states of the 2D compass model \cite{Doucot05} are coupled by perpendicular
global $X$ and $Z$ string operators and the phase measured using the
interferometry scheme above would also yield a\ phase $-1$\textbf{\ }due to
the anti-commutation relations at the crossing spin. Yet the ground states
are not topologically ordered because deformed string operators do not
preserve the ground subspace. Since our anyonic interferometry can test all
possible braiding operations, we can unambiguously verify the topological
property of anyons.


Various imperfections will degrade the signature of anyonic statistics. The
string operators may have errors that excite unwanted anyons, and the
topological memory may not fully restore to the ground state after braiding.
In addition, the topological memory may have anyons from imperfect
initialization. If these anyons are enclosed by the braiding loops, they
will affect the phase factor associated with braiding. However, neither of
these imperfections will prevent us from probing anyonic statistics, since
they only reduce the contrast of the anyonic interferometry without shifting
the fringes of $\left\langle \sigma_{\phi}\right\rangle $. We may even
distinguish the two types of imperfections from the contrast that depends on
different loops. The reduction of the contrast is proportional to the length
of the loops for errors from string operators, while it is proportional to
the area enclosed by the loops for errors from imperfect initialization (see
discussion in the Methods section).

\section{Probing and Control Anyonic Dynamics}

Our anyonic interferometry provides a tool to study the dynamics of anyons.
First, consider repeating the protocol [equation~(\ref{braiding3})] for
anyonic interferometry with the time delays $\left\{ t_{j}\right\} _{j=1,2,3}
$ between the four controlled-string operations. Processes of anyonic
creation, propagation, braiding, and annihilation will induce a time
dependence of the final state wavefunction in a general expression: $%
\left\vert \Psi _{final}\right\rangle =\alpha (\left\{ t_{j}\right\}
)\left\vert \Psi _{initial}\right\rangle +\beta (\left\{ t_{j}\right\}
)\left\vert \Psi _{\perp }(\left\{ t_{j}\right\} )\right\rangle $, where $%
\left\langle \Psi _{initial}|\Psi _{\perp }(\left\{ t_{j}\right\}
)\right\rangle =0$. Since the reduced density matrix of the probe qubit is $%
\rho =\frac{1}{2}\left(
\begin{array}{cc}
1 & \alpha (\left\{ t_{j}\right\} ) \\
\alpha ^{\ast }(\left\{ t_{j}\right\} ) & 1%
\end{array}%
\right) $, we can measure the complex coefficient $\alpha (\{t_{j}\})$ using
quantum state tomography \cite{NC00} of the probe qubit.

\begin{figure}[tbp]
\begin{center}
\includegraphics[width=8.7cm]{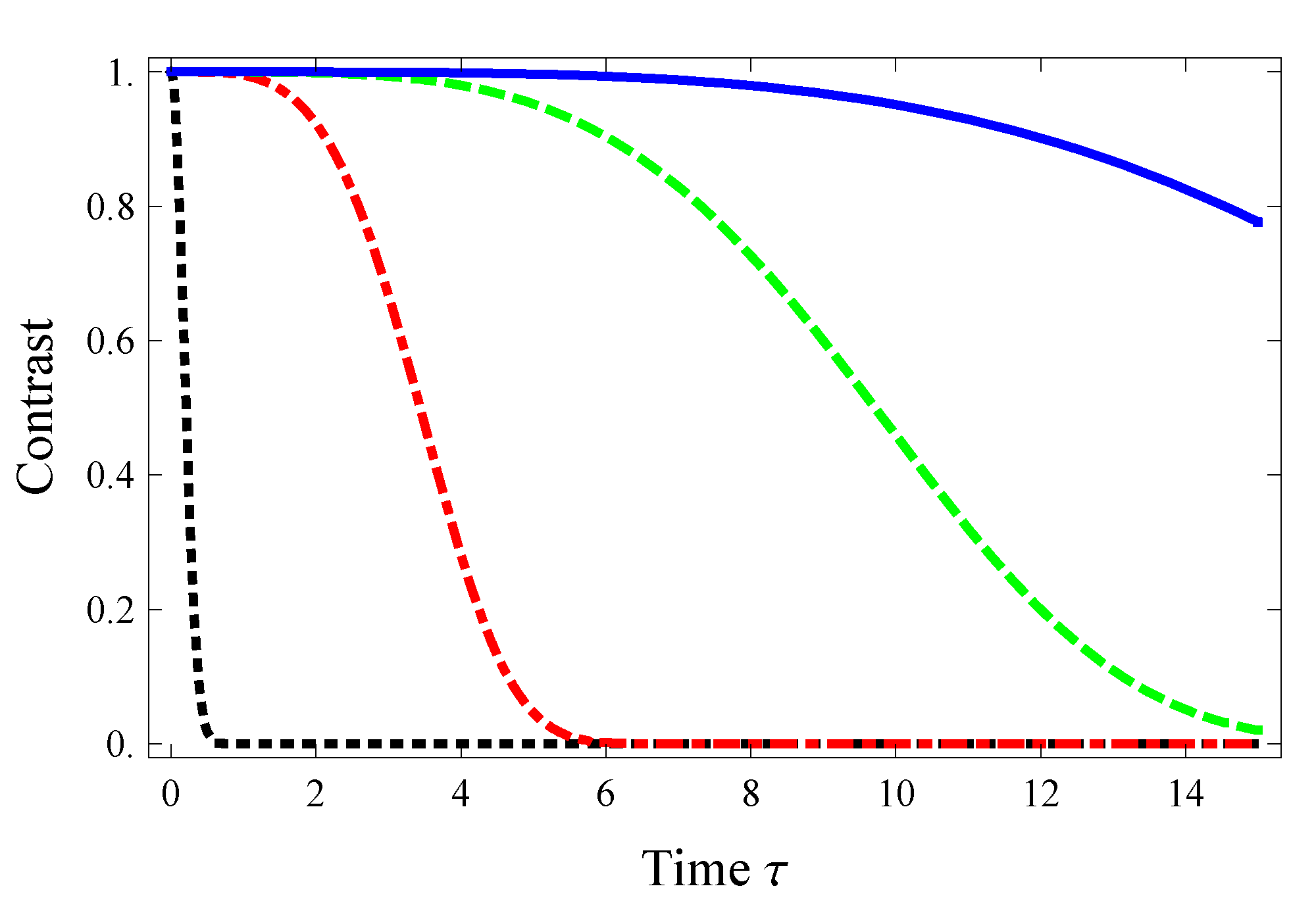}
\end{center}
\caption[fig:Contrast]{Fringe contrast of anyonic interferometry as a
function of the time for anyonic diffusion. The fringe contrast quickly
reduces due to anyonic diffusion (black dotted line). However, we can extend
the fringe contrast to longer times by applying one (red dotdashed line),
four (green dashed line), or ten (blue solid line) pairs of time-reversal
operations of $U_{\protect\pi }^{z}$ within the time interval $\protect\tau $%
. For clarity, we only consider the diffusion of two intermediate
x-particles induced by the perturbation $h_{e}^{x}\protect\sigma _{e}^{x}$
from equation~(\protect\ref{eq:HpertMain}). We choose the unit of time so
that the perturbation strength is normalized $\protect\sqrt{\left\langle
\left( h_{e}^{x}\right) ^{2}\right\rangle }=1$, and we assume that the
random field $h_{e}^{x}$ is a Gaussian random process with correlation time $%
\protect\tau _{c}=10$. (See SOM\ for details.)}
\label{fig:Contrast}
\end{figure}

Probing anyonic statistics can be regarded as special cases, with $\alpha
=e^{i\left( \theta +\eta \right) }$ or $e^{i\eta }$, and $\beta =0$.
Although the anyons are immobile for the surface code Hamiltonian, the
mobility of quasi-particles may change when we include local perturbations,
because the excited states with anyons are highly degenerate and any small
perturbation to the Hamiltonian can dramatically change the
eigen-wavefunctions. Consider for example a specific diffusion model for
anyons induced by the local perturbation
\begin{equation}
H_{pert}=\sum\limits_{\alpha \in \left\{ x,y,z\right\} }\sum\limits_{e\in
\mathrm{All~spins}}h_{e}^{\alpha }\sigma _{e}^{\alpha },
\label{eq:HpertMain}
\end{equation}%
where the $h_{e}^{x,z}$ field components cause diffusion of x(z)-particles
and the $h_{e}^{y}$ field component causes diffusion of dyonic particles
(pairs of neighboring x-particle and z-particle). (See SOM\ for details).
The nature of the perturbation (e.g., time independent or changing with
time) determines diffusion dynamics of anyons, which can be observed from
the coefficient $\alpha (\{t_{j}\})$ using our anyonic interferometry.

In addition, we can even control the diffusion dynamics of anyons. We
introduce the effective time-reversal operations $U_{\pi }^{z}\equiv
\prod_{e\in \text{All spins}}\sigma _{e}^{z}$ and $U_{\pi }^{x}\equiv
\prod_{e\in \text{All spins}}\sigma _{e}^{x}$, which anti-commute with $%
h_{e}^{x}\sigma _{e}^{x}$ and $h_{e}^{z}\sigma _{e}^{z}$ terms of $H_{pert}$%
, respectively. The combination of these operations (e.g., $U_{\pi }^{z}%
\overset{\tau /4}{\cdots \cdots }U_{\pi }^{x}\overset{\tau /4}{\cdots \cdots
}U_{\pi }^{z}\overset{\tau /4}{\cdots \cdots }U_{\pi }^{x}\overset{\tau /4}{%
\cdots \cdots }$) is analogous to spin-echo pulses in NMR, which can
effectively reverse anyonic diffusion caused by static perturbations and
consequently extend the fringe contrast of anyonic interferometry to longer
time delays, as illustrated in Fig.~\ref{fig:Contrast}. In essence, by
applying these global operations, we can localize the anyons without any
trapping potential. (See SOM\ for details.) Note that the anyonic
interferometry is closely related to the Ramsey experiments in atomic
physics \cite{Ramsey90}, which can now be performed with anyonic
quasi-particles.

\section{Outlook}

Controlled-string operations can be applied to other lattice Hamiltonians as
well \cite{Bacon06,Milman07}, which may provide robust quantum memory with
their degenerate ground states. For example subsystem codes \cite{Bacon06}
can be constructed out of 2D and 3D nearest neighbor spin-1/2 interactions
that are realizable with atomic systems \cite{Duan03,Micheli06}. Our
approach can be adapted to perform the logical operations generated by
strings or planes of Pauli operators in the 2D and 3D subsystem codes,
respectively. In addition, the ability to measure global operators on a spin
lattice provides a means to probe other properties of topological phases.
For example, a class of topologically ordered spin states known as string
net states \cite{Levin05}, which includes the ground states of $H_{\mathrm{%
surf}}$, have the property that they are invariant under large closed loop
operations. In the present case, these operators are $X_{\mathrm{loop}}(Z_{%
\mathrm{loop}})=\prod_{j\in C_{X,Z}^{closed}}\sigma ^{x,z}$ which have
expectation value $1$. A perturbation on $H_{\mathrm{surf}}$ in the regime $%
U\gg J$, by, e.g. a magnetic field, acts like a string tension that reduces
the amplitude of large loops (on a vacuum reference state). In fact there
are two phases as a function of the strength of the perturbation. For very
weak perturbations its has been argued that the loop order parameter scales
with the perimeter of the loop while for strong perturbations it scales with
the area \cite{Hastings05}. These are known as deconfined and confined
phases in analogy to lattice gauge theory and are examples of phenomena that
could be observed using our protocol. It may also be interesting to consider
adapting the present protocol to spin-lattice systems with non-abelian
anyons \cite{Aguado07}.%
%

\section{METHODS}

\subsection{Derivation of the geometric phase gate}

We describe the necessary elements to construct the geometric phase gate
illustrated in Fig.~\ref{fig:GeoPhaseGate}. First we require the
displacement operator $D(\xi)\equiv e^{\xi a^{\dagger}-\xi^{\ast}a}$ that
can be obtained by injecting coherent states through cavity mirrors. The
amplitude and phase of the injected field determine the phase space
displacement of the bosonic field by $\xi$.

Second, we need the displacement operation that depends on the state of the
memory spins:%
\begin{equation*}
D\left( \alpha e^{i\frac{\pi}{2}S_{\mathcal{C}}^{z}}\right) =\left\{
\begin{tabular}{cc}
$D\left( i\alpha\right) $ & if $\langle S_{\mathcal{C}}^{z}\rangle=+1$ \\
$D\left( -i\alpha\right) $ & if $\langle S_{\mathcal{C}}^{z}\rangle=-1$%
\end{tabular}
\ \ \ \ \ \right. ,
\end{equation*}
where we use $\langle S_{\mathcal{C}}^{z}\rangle=\pm1$ to represent the $%
\pm1 $ subspaces of the memory spins associated with the operator $S_{%
\mathcal{C}}^{z}$.
We can achieve $D\left( \alpha e^{i\frac{\pi}{2}S_{\mathcal{C}}^{z}}\right) $
by applying the QND Hamiltonian for time $t_{C}=\pi/2\chi$ both before and
after the displacement operation $D\left( \alpha e^{i\phi}\right) $. The
justification is based on the identity
\begin{equation*}
D\left( \alpha e^{i\phi+i\theta\Lambda}\right) =R\left( \theta O\right)
D\left( \alpha e^{i\phi}\right) R\left( -\theta O\right) ,
\end{equation*}
with $R\left( x\right) =e^{ixa^{\dag}a}$ and the two commuting operators $%
\left[ O,a\right] =0$. For $O=\sum_{j\in C}\sigma_{j}^{z}$ and $\theta
=\pi/2 $, we have $e^{i\phi+i\theta O}=e^{i\phi}\prod_{j\in C}\exp\left[ i%
\frac{\pi}{2}\sigma_{j}^{z}\right] =e^{i\left( \phi-m\pi/2\right)
}\prod_{j\in C}\sigma_{j}^{z}$, with $m=N_{C}$. Therefore, by choosing $%
\phi=\left( m+1\right) \pi/2$, we obtain $D\left( \alpha e^{i\theta
O+\phi}\right) =D\left( \alpha e^{i\frac{\pi}{2}S_{\mathcal{C}}^{z}}\right)
. $

Third, we need dispersive coupling between the bosonic field and the ancilla
spin (with two levels $\left\{ \left\vert 0\right\rangle _{A},\left\vert
1\right\rangle _{A}\right\} $)%
\begin{equation*}
V_{A}=\chi_{A}a^{\dag}a\left\vert 1\right\rangle _{A}\left\langle
1\right\vert ,
\end{equation*}
with coupling strength $\chi_{A}$, which can be switched on and off via
optical control \cite{Cho07,Gorshkov07} or mechanical displacement of the
ancilla spin. With such dispersive interaction, we are able to obtain the
displacement operation conditioned on the state of the ancilla spin, $%
D\left( \beta\left\vert 1\right\rangle _{A}\left\langle 1\right\vert \right)
=\left\vert 0\right\rangle _{A}\left\langle 0\right\vert \mathbf{\otimes I}%
+\left\vert 1\right\rangle _{A}\left\langle 1\right\vert \otimes D\left(
\beta\right) $, by the following procedure: (1) apply the interaction $V_{A}$
for time $t_{A}=\pi/\chi_{A}$, (2) displace the bosonic field by $-\beta/2$,
(3) apply the interaction $V_{A}$ again for time $t_{A}$, and (4) displace
the bosonic field by $\beta/2$. The steps (1-3) displace the bosonic field
by $\mp\beta/2$ for the ancilla spin in state $\left\vert 0\right\rangle
_{A} $ and $\left\vert 1\right\rangle _{A}$, respectively. Combined with the
displacement $\beta/2$ from step (4), we have the operation $D\left(
\beta\left\vert 1\right\rangle _{A}\left\langle 1\right\vert \right) $.

Finally, the controlled-string operation is a combination of the above
elements:%
\begin{equation}
U=D\left( -\beta\left\vert 1\right\rangle _{A}\left\langle 1\right\vert
_{A}\right) D\left( -\alpha e^{i\frac{\pi}{2}S_{\mathcal{C}}^{z}}\right)
D\left( \beta\left\vert 1\right\rangle _{A}\left\langle 1\right\vert
_{A}\right) D\left( \alpha e^{i\frac{\pi}{2}S_{\mathcal{C}}^{z}}\right) .
\label{eq:GeoEvolution}
\end{equation}
The bosonic field is restored to its initial state, while accumulating a
phase depending on both the state of the ancilla spin and the value for the
string operator as illustrated in Fig.~\ref{fig:GeoPhaseGate}.

\subsection{Fidelity of controlled-string operations and topological memory}

To evaluate the advantage of using topological memory storage, we compare
the improvement of storing a qubit in spins prepared in the ground states of
$H_{\mathrm{surf}}$ (assumed at zero temperature) versus the decoherence
rate for encoding a qubit in a single spin. For long-time storage of quantum
memory, we expect to gain from the robustness of the topological memory can
significantly reduce the decoherence rate by a factor of $\left( \delta
h/J\right) ^{N}\ll 1$, where $\delta h$ is the magnitude of the noise
perturbation on individual spins and $N$ is the length of the minimal string
associated with the generators for encoded qubits \cite{Kitaev03}. The
controlled-string operations that are used to swap quantum information in
and out have errors including cavity decay, radiative decay, and non uniform
dispersive shifts in the QND interaction.

\subsubsection{Errors due to photon loss}

The photon loss is attributed to two physical processes: the spontaneous
decay with rate $\gamma $ for the optically excited state $\left\vert
e\right\rangle $ (Fig.~\ref{fig:TopoMem1}b), and the cavity loss with rate $%
\kappa $ during the QND\ interaction. For single photon approach, the
interaction time is $\tau =\pi /2\chi $ and the effective spontaneous decay
rate is suppressed to $\frac{g^{2}}{\Delta ^{2}}\gamma $ by having large
detuning $\Delta \gg g$ for each selected spin (Fig.~\ref{fig:TopoMem1}b).
Under the assumption that the selected spins decay independently, the total
probability for photon loss is%
\begin{equation*}
\kappa \tau +N\frac{g^{2}}{\Delta ^{2}}\gamma \tau \geq \mathbf{2}\pi \sqrt{%
N/P}\equiv P_{loss},
\end{equation*}%
where we define the Purcell factor $P\equiv g^{2}/\kappa \gamma $ \cite%
{Purcell46,Michler00}. And the minimum probability can be achieved by
choosing optimal detuning $\Delta =g\sqrt{N\gamma /\kappa }$. 

For geometric phase gate approach, we can choose $\left\vert \alpha
\right\vert ^{2}=\left\vert \beta \right\vert ^{2}=\pi /2$ so that the total
probability for photon loss is $P_{loss}\approx \left\vert \alpha
\right\vert ^{2}\left( \kappa \tau +N\frac{g^{2}}{\Delta ^{2}}\gamma \tau
\right) $ with $\tau \approx \pi \Delta /g^{2}$. Similar to single photon
approach, the probability $P_{loss}$ can be significantly reduced by having
a large Purcell factor $P>N$. For coherent states, we cannot identify photon
loss events unambiguously, but we can still characterize the errors
associated with the photon loss.

\subsubsection{The deviation of the QND\ interaction}

The dominant deviation of the QND\ interaction is from the fluctuations of
the coupling strength between the cavity mode and selected spins, described
by the following perturbation%
\begin{equation*}
\delta H=\chi a^{\dag }a\sum_{j\in C}\delta _{j}\sigma _{j}^{z},
\end{equation*}%
where $\delta _{j}$ is the relative deviation for the $j$th spin. In the
presence of cavity excitation, the implementation of the gate $U_{j}=\exp
\left( i\theta \sigma _{j}^{z}\right) $ on the $j$th spin could lead to the
gate $\tilde{U}_{j}=\exp \left[ i\left( 1+\delta _{j}\right) \theta \sigma
_{j}^{z}\right] $. We define the error by the operator norm $\varepsilon
_{j}\equiv \left\Vert \tilde{U}_{j}-U_{j}\right\Vert \approx \theta
\left\vert \delta _{j}\right\vert $ \cite{Bhatia97}. Thus we have the error
for the QND\ interaction, $P_{QND}\equiv \left\Vert \prod_{j\in C}\tilde{U}%
_{j}-\prod_{j\in C}U_{j}\right\Vert \leq \sum_{j}\varepsilon _{j}$.

This kind of error due to inhomogeneity can be compensated in two ways.
First one could add an optical potential to the system which is shaped to
equalize the couplings $\delta_{j}$ for all the spins along the
configuration path $\mathcal{C}$. A second option is to use composite pulse
sequences on the system in which case it has been shown that the error can
be reduced to $O\left( |\delta_{j}|^{k}\right) $ for $\forall j$ using $%
O(k^{3})$ pulses \cite{Brown04}. Therefore, $P_{QND}\sim
N_{C}\theta\left\vert \delta \right\vert ^{k}$ is effectively suppressed.

\subsubsection{Summary}

Combining all the decoherence mechanisms, the error probability for the swap
in- swap-out process with memory storage time $t$ is%
\begin{equation*}
p_{\text{topo-mem}}=\left( \delta h/J\right) ^{N} q t+4\lambda \sqrt{N/P}%
+N\varepsilon,
\end{equation*}
with the pre-factor $\lambda=\mathbf{2}\pi$ (and $\mathbf{4}\pi^{2}$) for
the single photon (and geometric phase gate) approach. Compared with the
storage error without topological encoding $p_{\text{ref-mem}}=q\times t$,
for storage time $t>\frac{4\lambda\sqrt{N/P}+N\varepsilon}{q}\approx \frac{%
4\lambda\sqrt{N/P}}{q}$, our topological memory outperforms storage in a
single quantum system.

We remark that for single photon approach, photon loss induces leakage
errors that can be detected without compromising the state stored in the
topological memory. Such detected errors can be overcome by repetition,
which is applicable to probabilistic operations such as entanglement
generations in quantum repeater \cite{Briegel98} and distributed quantum
computer \cite{JTSL07}. Detected errors can have a very high tolerable
threshold for deterministic quantum computation schemes \cite{Knill05b}.

\subsection{Fringe contrast of the interferometer in the presence of
excitations}

We refer to anyons left from the initialization as \emph{quenched anyons},
which can result in measurable effects to the phase measurement associated
with braiding [equation~(\ref{braiding3})]. To be specific, we will consider
the planar code, and assume that the probability to have one pair of initial
anyons is $p$ while neglecting the case with multiple pairs of anyons. If
the anyons are immobile (e.g., the braiding operation is much faster than
anyonic propagation), the contrast of the phase measurement only depends the
probability that the loop $l_{1}\cup l_{3}$ (or $l_{2}^{\prime }\cup
l_{4}^{\prime }$) (see Fig.~\ref{fig:AnyonBraiding}a) encloses an odd number
of initial x-particles (or z-particles). An extra phase $e^{i\theta }=-1$
will be accumulated from each loop satisfying this condition. Suppose the
loop $l_{1}\cup l_{3}$ (or $l_{2}^{\prime }\cup l_{4}^{\prime }$) encloses $%
m $ faces (or $m^{\prime }$ vertices). If one pair of initial x-particles is
uniformly distributed among $N^{2}\times \left( N^{2}-1\right) /2$ possible
configurations, the probability for accumulating an extra phase is $%
q_{m}\equiv \frac{2m\left( N^{2}-m\right) }{N^{2}\left( N^{2}-1\right) }$
for the loop $l_{1}\cup l_{3}$. The probability reaches a maximum value $%
\approx 1/2$ for $m\approx N^{2}/2$; meanwhile it vanishes for $m=0$ or $%
N^{2}$, which is achieved by $l_{1}=l_{3}$. Therefore, the contrast for the
fringes of $\left\langle \sigma _{\phi }\right\rangle $ v.s. $\phi $ will be
reduced to $1-p\times \left( q_{m}+q_{m^{\prime }}\right) $.

If the anyons are highly diffusive (e.g., random anyonic propagation is very
fast compared to the intervals between the control operations), we should
avoid adding any anyons by applying string operators that act within the
ground subspace of the topological memory. As shown in Fig.~\ref%
{fig:AnyonBraiding}c, we use generators of the encoded qubits associated
with strings $\left\{ l_{Z1},l_{X2}^{\prime },l_{Z3},l_{X4}^{\prime
}\right\} $ to implement the braiding operation. However, any quenched
anyons (if present) will quickly diffuse over the entire torus and
completely wash away the fringe of $\left\langle \sigma _{\phi
}\right\rangle $. Therefore, the remaining contrast with highly diffusive
anyons is $1-p$. Note that imperfect string operators may also reduce the
contrast, since they may introduce unwanted anyons to the topological memory
with probability approximately proportional to the length of the string.

\subsection{Extension to $\mathbb{Z}_{d}$ gauge theories}

This interferometric technique can be extended to measure abelian anyonic
statistics for any $\mathbb{Z}_{d}$ gauge theory by introducing the spin
lattice Hamiltonian with $d$ levels for each spin \cite{Bullock07}. One can
still use a probe qubit to measure the statistical phase via
controlled-string operations. The mutual statistical phase between a charge $%
a\in\mathbb{Z}_{d}$ and flux $b\in\mathbb{Z}_{d}$ associated with the
braiding operation is $\tilde{Z}_{\mathcal{C}_{Z}^{\prime}}^{-a}\tilde{X}_{%
\mathcal{C}_{X}^{\prime}}^{-b}\tilde{Z}_{\mathcal{C}_{Z}}^{a}\tilde{X}_{%
\mathcal{C}_{X}}^{b}=e^{i2\pi ab/d}$. Here the string operator $\tilde{Z}_{%
\mathcal{C}_{Z}}^{a}$ ($\tilde {X}_{\mathcal{C}_{X}}^{b}$) is a product of $%
Z^{a}$ ($X^{b}$) operators of all the spins on the string $\mathcal{C}_{Z}$ (%
$\mathcal{C}_{X}$), where $Z$ and $X$ are elements of the generalized Pauli
group. The operator $Z$ can be implemented by phasing pairs of spin states
at a time using the protocols in the main text for the appropriate duration
at each stage. Equivalently, one can choose field polarizations and
detunings so that only one of the $d$ levels is strongly coupled to the
cavity, then evolve that state for the appropriate time, and swap other
states in, evolve, and swap out again. A total of $d-1$ global gates suffice
to simulate $\tilde{Z}_{\mathcal{C}_{Z} }^{a}$. The same follows for the $X_{%
\mathcal{C}_{X}}^{b}$ operators, but one must perform a parallel Fourier
transform operator $F=\prod_{j\in \mathcal{C}_{X}}f_{j}$ first on all the
spins in the configuration, then implement $Z_{\mathcal{C}_{X}}^{b}$ then
apply $F^{-1}$.


\section{Acknowledgements}

We gratefully acknowledge conversations with H. P. Buchler, L. Ioffe, and A.
M. Rey. Work at Harvard is supported by NSF, ARO-MURI, CUA, and the Packard
Foundation. Work at Innsbruck is supported by the Austrian Science
Foundation, the EU under grants OLAQUI, SCALA, and the Institute for Quantum
Information.

\section{Competing financial interests}

The authors declare that they have no competing financial interests.


\newpage 
\begin{center}
{\Large Supplementary Online Materials }
%
\end{center}

\section{Noise Model for Toric-Code Hamiltonian}

The toric-code Hamiltonian for spins on the edges of $N\times N$ square
lattice%
\begin{equation}
H_{topo}=-J\sum_{s}A_{s}-J\sum_{p}B_{p},
\end{equation}
is a sum of stabilizer operators $A_{s}=\Pi_{j\in\text{star}\left( s\right)
}\sigma_{j}^{x}$ and $B_{p}=\Pi_{j\in\text{boundary}\left( p\right)
}\sigma_{j}^{z}$ associated with the site (vertex) $s$ and the plaquette
(face) $p$, respectively. And the coupling strength $J$ determines the
energy gap between the ground and excited states $\Delta\sim J$, which is
also the energy associated with the quasi-particle excitations. There are
two types of quasi-particles: (1) \emph{z-particles} that live on the
vertices of the lattice and (2)\emph{\ x-particles} that live on the
plaquette. The quasi-particles do not change from one type to the other
type, but there is a non-trivial topological phase associated with braiding
of two quasi-particles of different type. We generate and move these
quasi-particles by applying string operators, meanwhile during the interval
between the string operators the quasi-particles will evolve under the
toric-code Hamiltonian and various local perturbations from the environment.
In the absence of perturbations from the environment, the quasi-particles
are immobile. However, the mobility of quasi-particles changes when we
include local perturbations, because the excited states with quasi-particles
are highly degenerate and any small perturbation to the Hamiltonian can
change both the energy spectrum and the eigen-wavefunctions.

\subsection{Perturbation Hamiltonian}

In this section, we will consider a simple model that will induce diffusion
of quasi-particles of the toric-code Hamiltonian. We will consider the case
that the perturbation is small compared to the energy gap $\Delta$, so that
he number of quasi-particles is still conserved. However, we will show that
such small perturbation can lead to non-trivial dynamics in the manifold
with fixed number of quasi-particles. The perturbation is described by the
Hamiltonian%
\begin{equation}
H_{pert}=\sum_{e\in\text{All spins}}h_{e}\sigma_{e}^{x}  \label{eq:Hpert}
\end{equation}
where $e$ is the label for spins that we sum over. The time dependent
coefficient $h_{e}$ is the local field associated with spin $e$.

We can also write the perturbation Hamiltonian by summing over the
plaquettes and their surrounding edges:%
\begin{equation}
H_{pert}=\frac{1}{2}\sum_{p}\sum_{\eta\in\mathcal{N}}h_{p,\eta}\sigma_{p,%
\eta }^{x},
\end{equation}
where $p$ is the label for the plaquette, $\eta$ is the label for the
surrounding edges, $\mathcal{N}=\left\{ \left( 0,1\right) ,\left(
0,-1\right) ,\left( 1,0\right) ,\left( -1,0\right) \right\} $ is the set
that includes four edges around the plaquette as illustrated in Fig.~\ref%
{fig:PlaquetteSite}a. The combination of $\left( p,\eta\right) $ label the $%
\eta$-edge of plaquette $p$. We use $h_{p,\eta}$ and $\sigma_{p,\eta}^{x}$
to represent the local fluctuating field and the spin operator for edge $%
\left( p,\eta\right) $, respectively. Since each edge is shared by two
neighboring plaquettes, both $\left( p,\eta\right) $ and $\left(
p+\eta,-\eta\right) $ represent the same edge. By definition, we have $%
h_{p,\eta}\equiv h_{p+\eta,-\eta}$ and $\sigma_{p,\eta}^{x}\equiv
\sigma_{p+\eta,-\eta}^{x}$. The prefactor $1/2$ in Eq. (\ref{eq:Hpert})
accounts for the double counting of edges.

\begin{figure}[ptbh]
\begin{center}
\includegraphics[width=8cm]{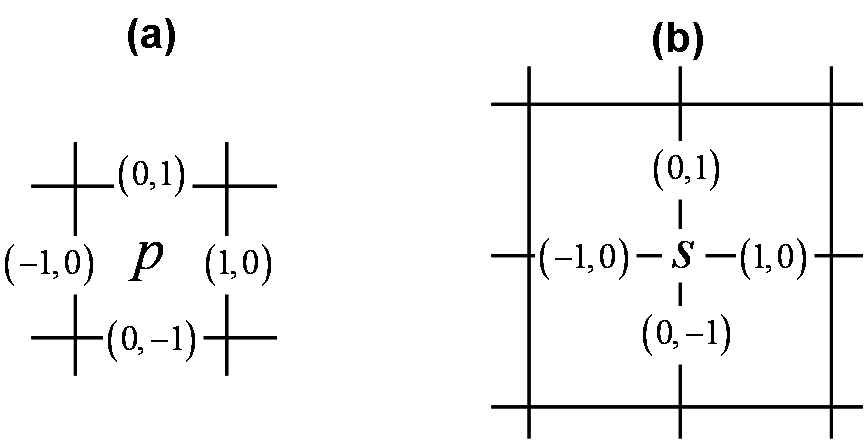}
\end{center}
\caption[fig:PlaquetteSite]{Relative position of edges with respect to (a)
plaquette and (b) site.}
\label{fig:PlaquetteSite}
\end{figure}

\subsection{Effects from Perturbation}

We now consider the effects from the perturbation Hamiltonian $H_{pert}$,
under the assumption that the number of quasi-particles is conserved. This
assumption can be justified as long as the local fluctuating field has a
small amplitude and low-frequency noise spectrum, compared with the energy
gap from the topological Hamiltonian. For ground states of the toric-code
Hamiltonian, there are no quasi-particle excitations. And the perturbation
Hamiltonian can only induce \emph{virtual} excitations of quasi-particles.
Thus the leading non-trivial effect within the ground-state manifold is the $%
N$th order perturbative process, associated with hopping of virtual
quasi-particles along a minimal non-contractible loop of length $N$. Such $N$%
th order process is suppressed by a factor $\left( \left\vert h\right\vert
/\Delta\right) ^{N}$ which decreases exponentially with respect to the
system size. As expected, the ground states of the topological Hamiltonian
should be robust against these local perturbations.

The perturbation Hamiltonian acts very differently on the excited states. It
can induce x-particles to hop to the neighboring plaquettes, while having no
influence to z-particles. We may define the creation operator of the
x-particle at plaquette $p$ as $b_{p}^{\dag}$, which changes the stabilizer $%
B_{p}$ from $+1$ to $-1$. Similarly, the annihilation operator $b_{p}$
changes $B_{p}$ from $-1$ to $+1$. By definition, $\left( b_{p}^{\dag
}\right) ^{2}=\left( b_{p}\right) ^{2}=0$. Applying $\sigma^{x}$ to one of
the edge of the plaquette will flip the sign of the stabilizer $B_{p}$,
represented by $b_{p}+b_{p}^{\dag}$. Since each edge is shared by two
neighboring plaquettes, $\sigma_{p,\eta}^{x}$ will flip both stabilizers $%
B_{p}$ and $B_{p+\eta,-\eta}$, which can be expressed as%
\begin{equation}
\sigma_{p,\eta}^{x}\rightarrow\left( b_{p}+b_{p}^{\dag}\right) \left(
b_{p+\eta}+b_{p+\eta}^{\dag}\right) .
\end{equation}
By conservation of quasi-particles, we can simplify the mapping:%
\begin{equation}
\sigma_{p,\eta}^{x}\rightarrow b_{p}b_{p+\eta}^{\dag}+b_{p}^{\dag}b_{p+\eta}.
\end{equation}
Therefore, within the manifold of fixed quasi-particles, the perturbation
Hamiltonian can be reduced to%
\begin{equation}
H_{pert}^{\prime}=\frac{1}{2}\sum_{p}\sum_{\eta\in\mathcal{N}}h_{p,\eta
}\left( b_{p}b_{p+\eta}^{\dag}+b_{p}^{\dag}b_{p+\eta}\right) ,
\end{equation}
where the local field $h_{p,\eta}$ can also be interpreted as the tunneling
rate of quasi-particles from $p$ to $p+\eta$, or from $p+\eta$ to $p$.

We now use $H_{pert}^{\prime}$ to study the dynamics of quasi-particles.
Suppose at time $t=0$ we create an x-particle at plaquette $p$%
\begin{equation}
\left\vert \varphi_{0}\right\rangle =b_{p}^{\dag}\left\vert vac\right\rangle
.
\end{equation}
And at a later time $\tau$ the state becomes%
\begin{align}
& \left\vert \varphi_{\tau}\right\rangle  \notag \\
& =\exp\left[ -iH_{pert}^{\prime}\tau\right] b_{p}^{\dag}\left\vert
vac\right\rangle  \notag \\
& =\left[ 1-\sum_{\eta}\left( h_{p,\eta}\tau\right) ^{2}\right]
b_{p}^{\dag}\left\vert vac\right\rangle +\sum_{\eta}\left(
h_{p,\eta}\tau\right) b_{p+\eta}^{\dag}\left\vert vac\right\rangle  \notag \\
& +O\left( h^{3}\tau^{3}\right) ,
\end{align}
where in the second equality we assume that $h_{p,\eta}$ is \emph{%
time-independent} and expand only to the second order of $h_{p,\eta}\tau$.
The overlap between the initial and final states is%
\begin{equation}
\left\langle \varphi_{0}|\varphi_{\tau}\right\rangle \approx1-\sum_{\eta
}\left( h_{p,\eta}\tau\right) ^{2}\approx\exp\left[ -\sum_{\eta}h_{p,\eta
}^{2}\tau^{2}\right] \text{.}
\end{equation}
Therefore, for time-independent perturbation and relatively short waiting
time (i.e., $\tau\ll h_{p,\eta}^{-1}$), the probability for a quasi-particle
to remain at the same position decreases \emph{quadratically} with time.

Generally, the local field $h_{p,\eta}$ will depend on time and we should
replace $h_{p,\eta}\tau$ by $\int_{0}^{\tau}h_{p,\eta}\left( t^{\prime
}\right) dt^{\prime}$ in the equations above. In particular, the local field
$h_{p,\eta}\left( t^{\prime}\right) $ can be a stochastic random variable.
In the next section, we will study the dynamics associated with stochastic
noise fields $\left\{ h_{p,\eta}\right\} $.

\subsection{Time-dependent Perturbation}

We now consider the local stochastic noise fields characterized by the
auto-correlation function
\begin{equation}
f\left( t\right) \equiv\overline{h_{p,\eta}\left( t^{\prime}\right)
h_{p,\eta}\left( t^{\prime}+t\right) },
\end{equation}
where we assume that $h_{p,\eta}$ is steady and the auto-correlation only
depends on the time difference between the two sampling points. For
simplicity, we will also assume independent local noise for different spins;
that is the correlation function $\overline{h_{p,\eta}\left( 0\right)
h_{p^{\prime},\eta^{\prime}}\left( t\right) }$ vanishes unless $\left(
p,\eta\right) $ and $\left( p^{\prime},\eta^{\prime}\right) $ represent the
same spin. We can characterize the noise by using the power spectrum
density, which is the Fourier transform of the auto-correlation function:
\begin{equation}
\tilde{f}\left( \omega\right) \equiv\frac{1}{2\pi}\int_{-\infty}^{\infty
}f\left( t\right) e^{-i\omega t}dt.
\end{equation}
For example, the Gaussian correlation function%
\begin{equation}
f\left( t\right) =\xi_{h}^{2}\exp\left[ -t^{2}/\tau_{c}^{2}\right]
\label{eq:GaussianCorr}
\end{equation}
has power spectrum density%
\begin{equation}
\tilde{f}\left( \omega\right) =\frac{\xi_{h}^{2}}{\sqrt{\pi}\omega_{c}}\exp%
\left[ -\omega^{2}/\omega_{c}^{2}\right] ,
\end{equation}
where the typical amplitude of the stochastic field is $\xi_{h}$, the
correlation time is $\tau_{c}$, and the cut-off frequency is $%
\omega_{c}=2\tau_{c}^{-1}$. The power spectrum density vanishes for high
frequency $\omega\gg\omega_{c}$.

We can estimate the probability for a quasi-particle to remain in the same
position, in the presence of local stochastic noise,%
\begin{align}
p_{\tau} & =\overline{\left\vert \left\langle \varphi_{0}|\varphi_{\tau
}\right\rangle \right\vert ^{2}}  \notag \\
& \approx1-2\sum_{\eta\in\mathcal{N}}\overline{\left( \int_{0}^{\tau
}h_{p,\eta}\left( t^{\prime}\right) dt^{\prime}\right) ^{2}}.
\label{eq:Prob}
\end{align}
We first evaluate the square average of the integral for the fluctuation
field%
\begin{align}
& \overline{\left( \int_{0}^{\tau}h_{p,\eta}\left( t^{\prime}\right)
dt^{\prime}\right) ^{2}}  \notag \\
& =\int_{0}^{\tau}\int_{0}^{\tau}\overline{h_{p,\eta}\left( t^{\prime
}\right) h_{p,\eta}\left( t^{\prime\prime}\right) }dt^{\prime}dt^{\prime%
\prime}  \notag \\
& =\int_{-\infty}^{\infty}\frac{\sin^{2}\left( \omega\tau/2\right) }{\left(
\omega\tau/2\right) ^{2}}\tilde{f}\left( \omega\right) d\omega  \notag \\
& \approx2\pi\tilde{f}\left( \omega=0\right) \tau,  \label{eq:IntegralAvg}
\end{align}
where the last step assumes that we are interested in a time scale much
longer than the noise correlation time: $\tau\gg\tau_{c}$. Plugging Eq. (\ref%
{eq:IntegralAvg}) into Eq. (\ref{eq:Prob}), we get the probability%
\begin{equation}
p_{\tau}\approx\exp\left[ -2z\Gamma\tau\right] ,
\end{equation}
where $z=\left\vert \mathcal{N}\right\vert =4$ is the coordination number of
the square lattice, and $\Gamma=2\pi\tilde{f}\left( \omega=0\right) $ is the
diffusion rate to each neighboring position. For Gaussian correlation [Eq. (%
\ref{eq:GaussianCorr})], the diffusion rate is
\begin{equation}
\Gamma=2\sqrt{\pi}\xi_{h}^{2}/\omega_{c}.
\end{equation}

\subsection{Fringe Contrast for Interference Experiment}

We now consider how diffusion of quasi-particles affect the fringe contrast
of the anyonic interferometry. The signal of the anyonic interferometry is
attributed to the interference from path-ways with different braiding of
anyonic quasi-particles, which is achieved by using controlled-string
operations. During the intervals between the controlled-string operations,
the quasi-particles excited by the controlled-string operations will
diffuse, and the final state of the topological memory will have components
orthogonal to the initial state. Since the orthogonal components do not
contribute to the fringes, the contrast will be reduced.

The anyonic interferometry is analogous to the Ramsey experiment in the
following aspects. For both cases, we start with a superposition state $%
\left\vert 0\right\rangle +\left\vert 1\right\rangle $ for some two level
system. Then, we let the system evolve, and meanwhile it is also interacting
with the environment (e.g., cavity mode and selected spins, or external
magnetic field). Finally, we projectively measure the system in some basis $%
\left\vert 0\right\rangle \pm e^{\pm i\phi }\left\vert 1\right\rangle $. The
reduction of the measurement signal is attributed to various decoherence
processes. For Ramsey experiment, the dominant decoherence is induced by
fluctuations of the external magnetic field and it is characterized by the
dephasing time $T_{2}^{\ast }$. For anyonic interferometry, we can define a
similar dephasing time%
\begin{equation}
T_{2}^{\ast }=\frac{1}{z\Gamma }.
\end{equation}%
And the fringe contrast is equal to the averaged overlap function, which can
be expressed as%
\begin{equation}
\overline{\left\vert \left\langle \varphi _{0}|\varphi _{\tau }\right\rangle
\right\vert }\approx \exp \left[ -\tau /T_{2}^{\ast }\right] .
\end{equation}%
For stochastic noise with Gaussian correlation [Eq. (\ref{eq:GaussianCorr}%
)], we have $T_{2}^{\ast }=\omega _{c}/\left( 2z\sqrt{\pi }\xi
_{h}^{2}\right) $.

\subsection{Spin Echo Techniques}

Similar to NMR systems, we can also use spin-echo techniques to further
suppress the stochastic noise. The essence of spin-echo is to apply an
effective time-reversal operation in the mid of the evolution so that the
noises from the two intervals cancel each other. For anyonic interferometry,
the effective time-reversal operation for the perturbation $H_{pert}$ is%
\begin{equation}
U_{\pi}^{z}\equiv\prod_{e\in\text{All spins}}\sigma_{e}^{z}.
\end{equation}
This is because $\left\{ H_{pert},U_{\pi}^{z}\right\} =0$ and $\left[
H_{topo},U_{\pi}^{z}\right] =0$. For example, at time $t=0$ we create an
x-particle at plaquette $p$%
\begin{equation}
\left\vert \varphi_{0}\right\rangle =b_{p}^{\dag}\left\vert vac\right\rangle
,
\end{equation}
We apply $U_{\pi}^{z}$ operations at time $\tau/2$ and $\tau$. The final
state will be%
\begin{align}
& \left\vert \varphi_{\tau}^{echo}\right\rangle  \notag \\
& =U_{\pi}^{z}e^{-i\int_{\tau/2}^{\tau}H_{pert}^{\prime}\left( t^{\prime
}\right) dt^{\prime}}U_{\pi}^{z}e^{-i\int_{0}^{\tau/2}H_{pert}^{\prime
}\left( t^{\prime}\right) dt^{\prime}}b_{p}^{\dag}\left\vert vac\right\rangle
\notag \\
& =e^{-i\int_{0}^{\tau/2}\left[ H_{pert}^{\prime}\left( t^{\prime}\right)
-H_{pert}^{\prime}\left( \tau/2+t^{\prime}\right) \right] dt^{%
\prime}}b_{p}^{\dag}\left\vert vac\right\rangle .
\end{align}
For stochastic noise with Gaussian correlation [Eq. (\ref{eq:GaussianCorr}%
)], the averaged overlap function now becomes%
\begin{equation}
\overline{\left\vert \left\langle
\varphi_{0}|\varphi_{\tau}^{echo}\right\rangle \right\vert }\approx\exp\left[
-\left( \tau/T_{2}\right) ^{4}\right]
\end{equation}
where $T_{2}\sim\sqrt{\tau_{c}/\xi_{h}}$. Since $T_{2}\sim
T_{2}^{\ast}\times\left( \tau_{c}\xi_{h}\right) ^{3/2}$, for $%
\tau_{c}\xi_{h}>1$ we can extend the coherence time by using echo-techniques.

Furthermore, we may introduce $n$ pairs of time-reversal operations within
the time interval $\tau$ (e.g., the pulse sequence $\underset{\text{Repeat }n%
\text{ times}}{\underbrace{U_{\pi}^{z}\overset{\tau/2n}{\cdots\cdots}U_{\pi
}^{z}\overset{\tau/2n}{\cdots\cdots}}}$). And the time dependence of the
constrast becomes%
\begin{equation*}
\exp\left[ -\frac{1}{n^{3}}\left( \frac{\tau}{T_{2}}\right) ^{4}\right] .
\end{equation*}
The contrast reduction is further slow down by a factor of $n^{3/4}$ and
this is illustrated in Fig.~5 of the paper.

\subsection{General Perturbation Hamiltonian}

We now generalize the perturbation Hamiltonian by including stochastic
fields along z direction%
\begin{equation}
H_{pert,gen}=\sum_{e\in\text{All spins}}h_{e}^{x}\sigma_{e}^{x}+h_{e}^{y}%
\sigma_{e}^{y}+h_{e}^{z}\sigma_{e}^{z}.  \label{eq:Hpert2}
\end{equation}
We introduce the creation (and annihilation) operators $a_{s}^{\dag}$ (and $%
a_{s}$) for z-particles at site $s$. Suppose the edge $e$ connects two sites
$s$ and $s+\zeta$ and it is also shared by two plaquettes $p$ and $p+\eta$
(see Fig.~\ref{fig:PlaquetteSite}). Within the manifold of fixed number of
quasi-particles, we have the following mapping:%
\begin{equation}
\sigma_{e}^{x}\rightarrow b_{p}b_{p+\eta}^{\dag}+b_{p}^{\dag}b_{p+\eta}
\end{equation}%
\begin{equation}
\sigma_{e}^{z}\rightarrow a_{s}a_{s+\zeta}^{\dag}+a_{s}^{\dag}a_{s+\zeta}
\end{equation}
Thus, the $\sigma_{e}^{x}$ term leads to hopping of x-particles, while the $%
\sigma_{e}^{z}$ term leads to hopping of z-particles. The $\sigma_{e}^{y}$
term leads to hopping of \emph{dyonic particles} (paired x-particle and
z-particle sharing the same edge, e.g., $a_{s}^{\dag}b_{p}^{\dag}$)%
\begin{equation}
\sigma_{e}^{y}=i\left[ \sigma_{e}^{x},\sigma_{e}^{z}\right] \rightarrow i%
\left[ b_{p}b_{p+\eta}^{\dag}+b_{p}^{\dag}b_{p+\eta},a_{s}a_{s+\zeta}^{\dag
}+a_{s}^{\dag}a_{s+\zeta}\right] .
\end{equation}
And the effective operator for $\sigma_{e}^{y}$ is still Hermitian. Note
that $\sigma_{e}^{y}$ consists of terms like $i\left(
b_{p}b_{p+\eta}^{\dag}a_{s}a_{s+\zeta}^{\dag}-a_{s}a_{s+\zeta}^{%
\dag}b_{p}b_{p+\eta}^{\dag}\right) $, and the minus sign for terms with
different order is consistent with the phase associated with the braiding of
anyons.

The generalized effective Hamiltonian becomes%
\begin{align}
& H_{pert,gen}^{\prime}  \notag \\
& =\frac{1}{2}\sum_{p}\sum_{\eta\in\mathcal{N}}h_{p,\eta}^{x}\left(
b_{p}b_{p+\eta}^{\dag}+b_{p}^{\dag}b_{p+\eta}\right)  \notag \\
& +\frac{1}{2}\sum_{s}\sum_{\zeta\in\mathcal{N}}h_{s,\zeta}^{z}\left(
a_{s}a_{s+\zeta}^{\dag}+a_{s}^{\dag}a_{s+\zeta}\right) \\
& +\frac{1}{4}\sum_{e\in\text{All spins}}ih_{e}^{y}\left[ b_{p}b_{p+\eta
}^{\dag}+b_{p}^{\dag}b_{p+\eta}~,~a_{s}a_{s+\zeta}^{\dag}+a_{s}^{\dag
}a_{s+\zeta}\right] .  \notag
\end{align}
where the first term induces hopping of x-particles and the second term for
z-particles. These two terms commute with each other, except for the
situation when $\left( p,\eta\right) $ and $\left( s,\zeta\right) $
represents the same edge. The third term induces hopping of dyonic
particles. If we are studying diffusion property of quasi-particles that are
far apart, there is essentially no dyonic particles and we may safely
neglect the effect from the third term. Therefore, as long as the diffusion
does not induce braiding of quasi-particles, we can safely treat the
diffusion for x-particles and z-particles as independent processes.

We may also introduce the effective time-reversal operation $%
U_{\pi}^{x}\equiv\prod_{e\in\text{All spins}}\sigma_{e}^{x}$ for
perturbations of $\sum_{e\in\text{All spins}}h_{e}^{z}\sigma_{e}^{z}.$,
since they anti-commute $\left\{ h_{e}^{z}\sigma_{e}^{z},U_{\pi}^{x}\right\}
=0$. We can combine $U_{\pi}^{z}$ and $U_{\pi}^{x}$ in a nested fashion to
suppress the diffusion of \textbf{both} x--particles and z-particles:%
\begin{equation*}
U_{\pi}^{z}\left( \overset{\tau/4}{\cdots\cdots}U_{\pi}^{x}\overset{\tau /4}{%
\cdots\cdots}U_{\pi}^{x}\right) U_{\pi}^{z}\left( U_{\pi}^{x}\overset{\tau/4}%
{\cdots\cdots}U_{\pi}^{x}\overset{\tau/4}{\cdots\cdots }\right) .
\end{equation*}
which can be further simplified as:%
\begin{equation}
U_{\pi}^{z}\overset{\tau/4}{\cdots\cdots}U_{\pi}^{x}\overset{\tau/4}{%
\cdots\cdots}U_{\pi}^{z}\overset{\tau/4}{\cdots\cdots}U_{\pi}^{x}\overset{%
\tau/4}{\cdots\cdots}
\end{equation}
Note that since time reversal operations $U_{\pi}^{x}$ and $U_{\pi}^{z}$
also anti-commute with $\sum_{e\in\text{All spins}}h_{e}^{z}\sigma_{e}^{z}$,
the nest combination of the two also suppress the diffusion of dyonic
particles. Therefore, we are able to suppress diffusion induced by the
general perturbation Hamiltonian of Eq. (\ref{eq:Hpert2}) to higher order.

\subsection{Time Reversal Operations for Surface-Code Hamiltonian with
Boundaries}

We now consider the effective time-reversal operations for the surface-code
Hamiltonian with boundaries. For the planar code on a square lattice (see
Fig.~1a of the paper), at the left and right are "rough edges" where the
stabilizer operator $B_{p}$ is a product of three $\sigma^{x}$ spin
operators associated with each boundary plaquette, while at the top and
bottom are "smooth edges" where the stabilizer operator $A_{s}$ is a product
of three $\sigma^{z}$ spin operators associated with each boundary site. In
contrast to the stabilizers associated with interior sites and plaquettes,
these boundary stabilizers anti-commute with the previous echo unitary $%
U_{\pi}^{x}$ or $U_{\pi}^{z}$. Thus, we have to modify the echo unitaries,
so that they commute with all stabilizers.

Let us first consider the modification of $U_{\pi}^{z}$. We refer to the
boundary protruding edges in even rows as "even rough edges," and those in
odd row as "odd rough edges." We define $U_{\pi}^{x,e}\equiv\prod_{\substack{
e\in \text{All spins}  \\ e\not \in \text{Odd rough edges}}}\sigma_{e}^{x}$
and $U_{\pi}^{x,o}\equiv\prod_{\substack{ e\in\text{All spins}  \\ e\not \in
\text{Even rough edges}}}\sigma_{e}^{x}$, so that $U_{\pi}^{x,e}$ and $%
U_{\pi }^{x,o}$ act on even and odd boundary edges, respectively. After this
modification, both $U_{\pi}^{x,e}$ and $U_{\pi}^{x,o}$ commute with all
stabilizers, especially $B_{p}$ at the left and right "rough edges."
Similarly, we label "even/odd smooth edges" associated the columns for the
top and bottom "smooth edges," and modify $U_{\pi}^{x}$ into $%
U_{\pi}^{z,e}\equiv\prod_{\substack{ e\in\text{All spins}  \\ e\not \in
\text{Odd smooth edges}}}\sigma_{e}^{z}$ and $U_{\pi}^{z,o}\equiv\prod
_{\substack{ e\in\text{All spins}  \\ e\not \in \text{Even smooth edges}}}%
\sigma_{e}^{z}$, which commute with all stabilizers.

We introduce the sub-sequence%
\begin{equation*}
W^{\alpha,\beta}\left( \tau\right) \equiv U_{\pi}^{z,\alpha}\overset{\tau/4}{%
\cdots\cdots}U_{\pi}^{x,\beta}\overset{\tau/4}{\cdots\cdots}U_{\pi
}^{z,\alpha}\overset{\tau/4}{\cdots\cdots}U_{\pi}^{x,\beta}\overset{\tau /4}{%
\cdots\cdots},
\end{equation*}
for $\alpha,\beta=e$ or $o$. And finally the full echo sequence to suppress
anyonic diffusion for the surface-code Hamiltonian with boundaries is%
\begin{equation}
W^{e,e}\left( \tau/4\right) W^{e,o}\left( \tau/4\right) W^{o,e}\left(
\tau/4\right) W^{o,o}\left( \tau/4\right) .
\end{equation}

\subsection{Summary}

In summary, we have analyzed a simple noise model on top of the toric-code
Hamiltonian. We have found that this noise model can explain diffusion of
quasi-particles. For anyonic interferometry, the effect of quasi-particle
diffusion is analogous to the dephasing of the Ramsey experiment. Based on
this analogy, we have proposed a scheme to extend the spin-echo technique to
the topological memory, which will further suppress the diffusion of
quasi-particles.

\section{Universal Rotations on the Topological Memory}

We can achieve universal rotations of the encoded qubit stored in the
topological memory. For example, an arbitrary x-rotation $\tilde{X}%
_{\theta}=e^{i\theta\tilde{X}}$ on the topological memory can be achieved
via the gate teleportation circuit
\[
\Qcircuit  @C=1em @R=1em @!R {
& \lstick{\ket{+}}
  &\qw  &\ctrl{1}  &\gate{X_{\theta}}   &\meter \cwx[1]
  \\
& \lstick{\ket{\Psi}}
  &{/} \qw  &\targ  &\qw                &\gate{\tilde{X}}       &\qw
  &\rstick{\tilde{X}_{\theta} \ket{\Psi} } \\
}
\]
This circuit represents the following procedure: (1) use controlled-string
operation $\Lambda\left[ \tilde{X}\right] $ to entangle the probe qubit
(upper line) and the memory (lower line with a slash), (2) projectively
measure the probe qubit in a rotated basis, and (3) perform an encoded Pauli
$X$ gate over the topological memory conditioned on the measurement outcome.
Similarly, we can also implement arbitrary z-rotation $\tilde{Z}%
_{\theta}=e^{i\theta\tilde{Z}}$ on the topological memory via the gate
teleportation circuit
\[
\Qcircuit  @C=1em @R=1em @!R {
& \lstick{\ket{0}}
  &\qw      &\targ      &\gate{Z_{\theta}}  &\gate{H}   &\meter \cwx[1]
  \\
& \lstick{\ket{\Psi}}
  &{/} \qw  &\ctrl{-1}  &\qw                &\qw        &\gate{\tilde{Z}}       &\qw
  &\rstick{\tilde{Z}_{\theta} \ket{\Psi} } \\
}
\]
Since any rotation can be decomposed into a sequence of x- and z-rotations,
the above two circuits suffice for universal rotations.

The gate teleportation can be generalized to implement arbitrary unitaries
generated by string operators. For string operator $\tilde{S}$, the unitary
operation $\tilde{S}_{\theta }=e^{i\theta \tilde{S}}$ can be achieved via
the gate teleportation circuit
\[
\Qcircuit  @C=1em @R=1em @!R {
& \lstick{\ket{+}}
  &\qw  &\ctrl{1}  &\gate{X_{\theta}}   &\meter \cwx[1]
  \\
& \lstick{\ket{\Psi}}
  &{/} \qw  &\gate{\tilde{S}}  &\qw                &\gate{\tilde{S}}       &\qw
  &\rstick{\tilde{S}_{\theta} \ket{\Psi} } \\
}
\]

For the geometric phase gate scheme, we can actually implement rotations of
the encoded qubit without the probe qubit, e.g., x-rotation of the encoded
qubit can be decomposed as $e^{i\theta\tilde{X}}=D\left( -\beta\right)
D\left( -\alpha e^{i\frac{\pi}{2}\tilde{X}}\right) D\left( \beta\right)
D\left( \alpha e^{i\frac{\pi}{2}\tilde{X}}\right) $ by choosing $%
|\alpha\beta|=\theta$.


\end{document}